\documentclass[epj]{svjour}
\usepackage{graphics}
\usepackage{amsmath}
\usepackage{amsfonts}
\usepackage{amssymb}
\usepackage{widetext}
\usepackage{xcolor}
\usepackage{hyperref}

\begin{document}
\title{Friedel oscillations of one-dimensional correlated fermions from perturbation theory and density functional theory}
\author{J.~Odavi\'c\inst{1}, N.~Helbig\inst{1,2,3} \and V.~Meden\inst{1}
}                     
\institute{Institut f{\"u}r Theorie der Statistischen Physik, RWTH Aachen University and JARA---Fundamentals of Future Information Technology, 52056 Aachen, Germany \and Peter-Gr\"unberg Institut and Institute for Advanced Simulation, Forschungszentrum J\"ulich, D-52425 J\"ulich, Germany \and nanomat/QMAT/CESAM and Department of Physics, Universit\'e de Lie\`ge, 4000 Li\`ege, Belgium}
%
%
\abstract{
  We study the asymptotic decay of the Friedel density oscillations induced by an open
  boundary in a one-dimensional chain of lattice fermions with a short-range two-particle
  interaction. From Tomonaga-Luttinger liquid theory it is known that the decay follows a
  power law, with an interaction dependent exponent, which, for repulsive interactions, is
  larger than the noninteracting value $-1$. We first investigate if this behavior can be
  captured by many-body perturbation theory for either the Green function or the self-energy
  in lowest order in the two-particle interaction. The analytic results of the former show a
  logarithmic divergence indicative of the power law. One might hope that the resummation
  of higher order terms inherent to the Dyson equation then leads to a power law in the
  perturbation theory for the self-energy.
  However, the numerical results do not support this.
  Next we use density functional theory within the local-density approximation
  and an exchange-correlation functional derived from the exact Bethe ansatz solution of the
  translational invariant model. While the numerical results are consistent with power-law
  scaling if systems of $10^4$ or more lattice sites are considered, the extracted exponent is very
  close to the noninteracting value even for sizeable interactions. 
%
} 
\authorrunning{J.~Odavi\'c, N.~Helbig \and V.~Meden}
\titlerunning{Friedel oscillations of one-dimensional correlated fermions from PT and DFT}
\maketitle
\section{Introduction}\label{sec:intro}
The elementary excitations of one-dimensional (1d), metallic Fermi systems with a two-particle
interaction are not given by fermionic quasi-particles, but are instead of collective, bosonic
nature \cite{Schoenhammer05,Giamarchi03}.  Such quantum many-body systems can thus not be described
by Fermi liquid theory. For short-ranged, i.e.~screended, two-particle interactions, on which we focus
here, Tomonaga-Luttinger liquid theory is applicable instead \cite{Haldane81}. One of the
characteristics of Tomonaga-Luttinger liquids is the power-law decay of correlation functions
at large times or spatial distances with exponents which, in spinless models, can be
expressed in terms of a single parameter $K$. This Tomonaga-Luttinger liquid parameter depends
on the band structure and filling as well as on the amplitude and range of the two-particle
interaction of the model Hamiltonian. For repulsive
interactions $0<K<1$ while $K>1$ for attractive ones; $K=1$ corresponds to noninteracting fermions.

To exemplify the Tomonaga-Luttinger liquid behavior let us focus on the observable of interest
to us, which is the density $n(x)$. Depending on the model considered the spatial variable $x$
might be continuous or given by a lattice site index $x \to j=1,2,\ldots,L$ and $L$ being the
system size (the lattice spacing is set to 1). We consider a system with open boundary conditions
in which translational invariance is broken. Generically, $n(x)$ shows oscillations which decay
from the boundaries towards the middle, the bulk part of the chain, at which the
average density $\nu$ is reached. From Tomonaga-Luttinger liquid theory it is known
that $n(x)-\nu$ decays as $x^{-K}$ and oscillates with (spatial) frequency
$2 k_{\rm F}$, with the Fermi momentum $k_{\rm F}$  ($\hbar=1$) \cite{Fabrizio95,Egger95}.
For a single-band lattice model $k_{\rm F}=\nu \pi$. These are the famous
Friedel oscillations with an exponent which, however, is modified by the interaction
as compared to the noninteracting value $-1$ ($-d$ in $d$ dimensions). For repulsive interactions
the oscillations decay slower while they decay faster for attractive ones.

For the lattice model of spinless fermions with nearest-neighbor hopping $t$ and nearest-neighbor
density-density interaction $U$ considered here, in the thermodynamic limit
$K(\nu,U/t)$ can be expressed in terms of a set
of coupled integral equations derived from the Bethe ansatz solution of this
model \cite{Haldane80}. At half-filling, $\nu=1/2$, a closed-form expression for $K(\nu,U/t)$ can
be derived. For $-2 < U/t<2$ the model is in a metallic Tomonaga-Luttinger liquid phase while
for $|U|/t > 2$ insulating phases are found. Away from half-filling the model is a
Tomonaga-Luttinger liquid for all $U/t>-2$. However, the integral equations can only be solved
numerically (with high precision) and accordingly $K(\nu,U/t)$ is only known numerically. We
will refer to this as the exact Tomonaga-Luttinger liquid parameter.    

It is generally believed that approximate approaches to the quantum many-body problem which
lead to an effective fermionic single-particle picture, such as, e.g., lowest order
perturbation theory, will generically fail to capture Tomonaga-Luttinger liquid behavior
of correlation functions. Such approaches appear to be at odds with the absence of
fermionic quasi-particles in Tomonaga-Luttinger liquids.
An exception to this is the local single-particle spectral function as a function of
frequency on lattice sites close to an
open boundary. For this the lowest order perturbation theory in $U$ for the self-energy,
i.e.~the non-self-consistent Hartree-Fock approximation, leads to a power-law suppression
in accordance with Tomonaga-Luttinger liquid theory \cite{Meden00}.  Motivated by this we
investigate if the same holds for the decay of the density oscillations away from an
open boundary and into the bulk of the chain.
Lowest order perturbation theory for the Green function shows a logarithmic position dependence
consistent with the power law. However, the numerical non-self-consistent Hartree-Fock data do
not support that the resummation of higher-order terms inherent to the Dyson equation 
does elevate this logarithmic term to a power law.
 
Next, we study if the power-law decay of the density with an interaction
dependent exponent can be obtained within (lattice \cite{Gunnarsson86,Schoenhammer95,Lima03,Schmitteckert_Evers_2008})
density functional theory (DFT) \cite{Hohenberg64}, an approach which also
builds on an effective single-particle picture. We employ the local density
approximation extracted from the ground-state energy obtained from the Bethe ansatz (BALDA)
solution of the lattice model.  For a different lattice model this approach was first suggested in Ref.~\cite{Schoenhammer95}.  

In Ref.~\cite{Schenk08} BALDA-DFT was used to investigate the static and dynamic
response of the translational invariant (periodic boundary conditions)
lattice model described above as well as the behavior of this model if
a single impurity is introduced. The authors concluded that 
Tomonaga-Luttinger liquid behavior is not captured. However, they did not
search for the characteristic power-law scaling of correlation functions and were
bound to systems of only a few hundred lattice sites (see below).

The two observables which are most directly accessible
within a DFT approach are the ground-state energy and the ground-state density. Here, we study
the latter for systems of up to $10^6$ lattice sites; the former does not contain any Tomonaga-Luttinger
liquid power laws \cite{Schoenhammer05,Giamarchi03}. The characteristic Tomonaga-Luttinger liquid
behavior induced by an open boundary is much less involved than the one resulting from a localized
impurity (renormalization group flow of the impurity towards an open boundary)
\cite{Luther74,Apel82,Kane92,Egger95}. Posing the question if BALDA-DFT can correctly
describe the decay of the Friedel oscillations due to an open boundary, as we do here, thus
constitutes less of a challenge to this method as compared to the
problems investigated in Ref.~\cite{Schenk08}. With a different emphasize to ours the perspectives of using Hartree-Fock and DFT to study Friedel oscillations in 1d correlated fermions were also investigated in Ref.~\cite{Sch}

With the hard wall boundary replaced by a local impurity
Friedel oscillations were investigated for the 1d (spinful) Hubbard model
in Ref.~\cite{Lima03} employing BALDA-DFT. The density for lattices of a few
hundred lattice sites was computed. For the local single-particle spectral function 
it is well established that to unambiguously observe asymptotic Tomonaga-Luttinger
liquid power-law behavior much larger system sizes (of the order of $10^4$ to $10^5$ sites) are
required even in the most simple case of spinless fermions with open boundaries; see
e.g.~Ref.~\cite{Andergassen04}. The same is expected to hold for the decay of the density
oscillations; see below for explicit results on this. For smaller systems the
asymptotics is completely masked by finite size effects. The study of Ref.~\cite{Lima03}
faces two additional challenges: (1) Due to the logarithmically slow vanishing of the
two-particle backscattering of particles with opposite spin for increasing system
size \cite{Solyom79}, even larger systems than for spinless models are required to observe
power laws in the Hubbard model, see e.g.~Refs.~\cite{Andergassen06,Soeffing13}. (2) The
finite local impurity of Ref.~\cite{Lima03} requires larger systems to observe the
asymptotic density decay than the open boundary, see
e.g.~Refs.~\cite{Kane92,Egger95,Andergassen04,Andergassen06}. It was thus premature to fit the
density decay obtained in Ref.~\cite{Lima03} by a power law.
Not surprisingly, the exponents obtained for repulsive interactions
are smaller than $-1$ and, therefore,
contradict Tomonaga-Luttinger liquid theory. 

Our numerical BALDA-DFT data for the density decay away from the open boundary for the above
lattice model of spinless fermions turn out to be consistent with power-law
scaling if system sizes of $10^4$ or more lattice sites are considered. However, the exponent
extracted is very different from the exact Tomonaga-Luttinger liquid parameter. Even for sizeable
interactions the data appear to be consistent with the noninteracting value $-1$.

The remainder of this paper is organized as follows. In Sect.~\ref{sec:modelmethods} we
present our model and give basics on the methods used to compute the density. Our
results obtained by the three approaches, i.e.~lowest order perturbation theory for the Green function,
the non-self-consistent Hartree-Fock approximation as well as the BALDA-DFT, are presented
in the three subsections of Sect.~\ref{sec:results}. Details on the analytical calculations for
the Green function perturbation theory are given in the Appendix. We conclude in Sect.~\ref{sec:conclusion}.

\section{The model and methods}
\label{sec:modelmethods}

\subsection{Spinless lattice fermions}
\label{sec:spinlferm}

We study the 1d model of spinless fermions with nearest-neighbor hopping $t>0$ and
nearest-neighbor interaction $U$ between particles occupying the Wannier states
with lattice site index $j$. It is given by the Hamiltonian
\begin{eqnarray}
  \label{eq:ham}
  H = - t \sum_{j=1}^{L-1} \left( c_{j+1}^\dag c_j^{\phantom{\dag}} + \mbox{H.c.} \right)
  + U \sum_{j=1}^{L-1} n_j n_{j+1}
\end{eqnarray}
in standard second quantized notation,
where $n_j = c_j^\dag c_j^{\phantom{\dag}}$ is the density operator on site $j$ and $L$
denotes the number of lattice sites. Note the open boundary conditions. 

In the noninteracting case, $U=0$, the single-particle
eigenfunctions $\left| n \right> $, with $n \in \{ 1,2,...,L \}$ are given by
\begin{equation}
  \label{eq:U0es}
  \langle j \vert n \rangle =
  \sqrt{\frac{2}{L+1}} \sin{(k_{n} j)}, \quad k_{n} = \frac{n \pi}{(L+1)}.
\end{equation}
The single-particle energies are $\epsilon(k) = - 2 t \cos{k}$.  
The many-body ground state for band filling $\nu = N/L$ is given by the Slater determinant
build out of the first $N$ single-particle states.

The ground-state expectation value of the density can, for arbitrary $U$, be computed from
the (zero temperature) Matsubara Green function $G_{j,j'}(\omega)$  as 
\begin{equation}
  \label{eq:densGreen}
  n(j) = \langle n_j \rangle =
  \frac{1}{2} + \frac{1}{\pi} \int \limits_{0}^{\infty} d \omega \, {\rm Re}\, G_{j,j}(i \omega) . 
\end{equation}

For $U=0$ the Green function in the single-particle eigenbasis $\left\{\left| n \right> \right\}$
is given by
\begin{equation}
  \label{eq:Green0}
  G^{0}_{n,n'} = [i \omega  - \xi (k_{n})]^{-1} \delta_{n,n'},
\end{equation}
where
$\xi (k)=\epsilon (k) - \mu$, with the chemical potential $\mu$.  
Changing to this basis and inserting $G^0$, the integral in Eq.~(\ref{eq:densGreen}) can be performed
leading to ($j=1,2,\ldots,L$)
\begin{equation}
  \label{eq:n0}
  n^{0}(j) = \frac{2 N +1}{2 (L+1)} - \frac{1}{2 (L+1)}
  \frac{\sin{\left( \frac{\pi}{L+1} j \left[ 2N+1 \right] \right)}}{\sin{\left( \frac{\pi}{L+1} j
      \right)}} .
\end{equation}  
In the thermodynamic limit $L,N \rightarrow \infty$, $\nu = N/L$ fixed, the noninteracting
density reduces to
\begin{equation}
\label{eq:n0TD}
  n^{0}(j) = \nu - \frac{\sin{( 2 k_{{\rm F}} j)}}{2 \pi j} . 
\end{equation}
These are the well known Friedel oscillations with wave vector $2 k_{\rm F}$ which, in a
noninteracting 1d system, decay as $1/j$.

Note that for half filling, $\nu= 1/2$, of the lattice the oscillatory part of
the density vanishes in both the finite system result
Eq.~(\ref{eq:n0}) as well as in the $L \to \infty$ result Eq.~(\ref{eq:n0TD}) and
$n^{0}(j) = 1/2$. The same holds for $U \neq 0$ \cite{Kitanine08}. 

\subsection{The Bethe ansatz solution and Tomonaga-Luttinger liquid properties}

For $U \neq 0$ the model Eq.~(\ref{eq:ham}) with periodic boundary conditions is
Bethe ansatz solvable (see e.g.~Ref.~\cite{Giamarchi03}).
In the thermodynamic limit this allows one to formulate a closed set of integral equations
from which the ground-state energy and other quantities of interest can be
obtained. The results derived along this line are consistent with the assumption
that the model falls into the Tomonaga-Luttinger liquid universality class for
$\nu \neq 1/2$ and all $U/t>-2$ as well as for $-2 < U/t < 2$ at
half filling $\nu=1/2$ \cite{Haldane80}. We focus on this Tomonaga-Luttinger liquid regime.  

The solvability by Bethe ansatz does, however, not imply that explicit analytic
expressions for correlation functions showing the characteristic Tomonaga-Luttinger
liquid power laws can be derived. Computing correlation functions by numerical methods
(for particularly convincing results, see Ref.~\cite{Karrasch12}) as well as
renormalization group approaches (see e.g.~Refs.~\cite{Giamarchi03,Solyom79,Andergassen04,Schmitteckert_Eckern_1996}) it was
still unambiguously confirmed that, for the above parameter regime, the model is a
Tomonaga-Luttinger liquid. The corresponding asymptotic decay of the Friedel
oscillations off an open boundary
\begin{equation}
  \label{eq:friedeldecayU}
  \left|n^{\rm TL}(j) - \nu \right| \sim  \frac{\sin {( 2 k_{{\rm F}} j)}}{j^K} ,
\end{equation}   
as described in Sect.~\ref{sec:intro}, was explicitely confirmed for the present
model in Ref.~\cite{Andergassen04}.

For $\nu \neq 1/2$ results for the Tomonaga-Luttinger liquid
parameter $K(\nu,U/t)$ can be obtained from numerically solving the Bethe ansatz
integral equations \cite{Haldane80}. To leading order in $U/t$ one
finds \cite{Meden00,Giamarchi03}
\begin{equation}
  \label{eq:KsmallU}
  K=1- \frac{U}{\pi v_{\rm F}} \left[ 1 - \cos \left( 2 k_{\rm F} \right) \right] + {\mathcal O}\left( [U/t]^2
    \right) ,
\end{equation}
with the Fermi velocity $v_{\rm F} = 2 t \sin k_{\rm F}$. 
For $\nu=1/2$ a closed analytical expression for $K(1/2,U/t)$ can be derived even beyond the leading
order. However, as already indicated by the absence of Friedel oscillations for $U=0$ [see Eqs.~(\ref{eq:n0})
and (\ref{eq:n0TD})], half-filling is nongeneric if it comes to density oscillations and
thus of minor interest to us.

\subsection{The Bethe ansatz solution and LDA-DFT}

The Bethe ansatz integral equations for the translational invariant model can also be used
within a Bethe ansatz (BA)LDA-DFT approach to derive an exchange-correlation functional.

In a practical implementation of the DFT idea one constructs an auxiliary,
noninteracting Kohn-Sham Hamiltonian \cite{Sham66}
\begin{equation}
  \label{eq:HKS}
H^{\rm KS} = - t \sum_{j=1}^{L-1} \left( c_{j+1}^\dag c_j^{\phantom{\dag}} + \mbox{H.c.} \right)
  + \sum_{j=1}^{L-1}   v_{j} n_j ,
\end{equation}  
with the onsite potential $v_{j}$ chosen such that it leads to the same density
$n(j)$ as in the interacting problem.
The single-particle potential is written as  $v_j = v_j^{\rm H} + v_j^{\rm xc}$
with the Hartree potential 
$v_j^{\rm H}  = U\left[ n(j+1) + n(j-1) \right]$ and the
exchange-correlation potential on site $j$
\begin{equation}
  \label{eq:Vxc}
  v_{j}^{{\rm xc}} =
  \frac{\partial }{\partial n}\Big[ e^{{\rm BA}} (n,U) - e^{{\rm H}}(n,U) \Big]_{n=n(j)}
\end{equation}
where $e^{{\rm BA}} (n,U)$ is the Bethe ansatz ground-state energy per site of the homogeneous
system with density $n$ and interaction strength $U$.  The other term is the
Hartree energy given by $e^{{\rm H}} (n,U)= - \frac{2 t}{\pi} \sin{(k_{{\rm F}})} + U n^2$.
The exchange-correlation potential
is computed numerically solving the Bethe ansatz integral equations.
The derivative in Eq.~(\ref{eq:Vxc}) is approximated by centered differences. For a plot of
$v^{{\rm xc}}$ as a function of the density at different $U/t$ for our model,
see Fig.~1 of Ref.~\cite{Schenk08}.

When numerically solving the DFT self-consistency problem, instead of following the standard procedure
of diagonalizing the single-particle Kohn-Sham Hamiltonian Eq.~(\ref{eq:HKS}) and
subsequently computing the density from the Kohn-Sham single-particle eigenstates,
we here proceed differently. To compute the $(j,j)$ matrix element of the Green function of
the Kohn-Sham system, from which $n^{\rm DFT}(j)$ can be obtained by Eq.~(\ref{eq:densGreen}), we
only have to determine the diagonal part of the inverse of the tri-diagonal matrix associated
to $H^{\rm KS}$. As described in Appendix C of Ref.~\cite{Andergassen04} this can be achieved in
${\mathcal O}(L)$ time ($L$ is the system size and thus the size of the resolvent matrix)
and is thus much faster and requires less memory as compared to a diagonalization.
We note that we implement the integration of Eq.~(\ref{eq:densGreen}) as the solution of
a differential equation (for more details see \cite{Odavic19}). All this allows us to study systems of up to $10^6$ lattice sites (at sizeable filling) not
accessible following the standard procedure. In particular, we are able to study
much larger systems as compared to the
ones investigated in Refs.~\cite{Schenk08} (spinless fermions) and \cite{Lima03}
(Hubbard model). We believe that this approach might also be useful in other DFT
applications. The self-consistency cycle of DFT was stopped when the change of the
density summed over all lattice sites was less than $10^{-5}$. The convergence
is achieved in about 10 to 20 cycles, when performing the usual linear mixing of
the density. 

Results for the density profile obtained along these lines are presented in Sect.~\ref{subsec:DFT}.

\subsection{Perturbation theory}
\label{sec:PT}

  Many-body perturbation theory in lowest order in $U/t$ provides an alternative way to obtain approximate
results for the density. We compute the self-energy $\Sigma$ to first order in $U/t$
(non-self-consistent Hartree-Fock approximation). To this order it becomes (Matsubara-) frequency
independent. Within the Wannier basis $\Sigma^{\rm 1PT}$ is a tri-diagonal matrix with the diagonal
(Hartree term) given by
\begin{equation}
  \label{eq:SigmaHFdia}
  \Sigma_{j,j}^{\rm 1PT}\! = \! - U \!\! \times \!\! \left\{ \begin{array}{ll}
                           \!\!  n^0(2)   & \!  \mbox{for} \, j=1 \\
                             \!\!  \left[ n^0(j-1) + n^0(j+1) \right] &  \! \mbox{for} \, j=2,\ldots,L-1 \\
                            \!\!    n^0(L-1)   &  \! \mbox{for} \, j=L ,
                           \end{array} \right. 
\end{equation}
with $n^0(j)$ stated in Eq.~(\ref{eq:n0}). The upper first off-diagonal (Fock term) reads

\begin{align}
  \label{eq:SigmaHFoffdia}
  \Sigma_{j,j+1}^{\rm 1PT} = & \frac{U}{2(L+1)} \left\{ \frac{\sin{\left[ \frac{\pi}{L+1} \left( N + \frac{1}{2}
                     \right)  \right]}}{\sin\left[ \frac{\pi}{2(L+1)} \right]}  \right. \nonumber \\
                   & - \left. \frac{\sin{\left[ \frac{\pi}{L+1} \left( N + \frac{1}{2}
                     \right) \left(2 j +1\right) \right]}}{\sin\left[ \frac{\pi}{2(L+1)} (2j+1) \right]} 
                     \right\},
\end{align}
with $j=1,2,\ldots,L-1$. The lower first off-diagonal follows from $\Sigma^\dag = \Sigma$.    
To obtain the non-self-consistent Hartree-Fock approximation for the Green function 
\begin{equation}
  \label{eq:Dyson}
  G^{\rm HF}= \left[ \left(G^0\right)^{-1} - \Sigma^{\rm 1PT} \right]^{-1}
\end{equation}
and from this $n^{\rm HF}(j)$ employing Eq.~(\ref{eq:densGreen}),
we thus have to solve a noninteracting single-particle problem with an
effective bond-dependent nearest-neighbor hopping $t-\Sigma^{\rm 1PT}_{j,j+1}$ and
the effective site-dependent onsite energy $\Sigma^{\rm 1PT}_{j,j}$. 
Due to the involved $j$ dependence of the hopping and onsite energy, reflecting
the Friedel oscillations of the noninteracting density $n^0(j)$, this cannot be
achieved analytically. As in BALDA-DFT we refrain from numerically
diagonalizing the effective single-particle Hamiltonian and instead exploit that to compute
$n^{\rm HF}(j)$ we only need the diagonal part of the inverse of a tri-diagonal
matrix which can be determined numerically in ${\mathcal O}(L)$ \cite{Andergassen04}.
For results, see Sect.~\ref{subsec:self}.

To gain analytical insights we expand Eq.~(\ref{eq:Dyson}) to first order in $U$
(first order perturbation theory for the Green function)
\begin{equation}
  \label{eq:Dysonexp}
  G^{\rm 1PT}= G^0 + G^0 \Sigma^{\rm 1PT} G^0
\end{equation}
which, using Eq.~(\ref{eq:densGreen}), leads to a first order approximation for the
density $n^{\rm 1PT}(j)$. For analytical calculations it is advantageous to work in
the basis of the single-particle
eigenfunctions $\left\{\left| n \right> \right\}$ of the
noninteracting Hamiltonian Eq.~(\ref{eq:U0es}) in which $G^0$ is diagonal; see Eq.~(\ref{eq:Green0}).
We thus have to compute $\Sigma^{\rm 1PT}$ in this basis instead of the Wannier basis as done in
Eqs.~(\ref{eq:SigmaHFdia}) and (\ref{eq:SigmaHFoffdia}). Details
on this and the corresponding results for $n^{\rm 1PT}(j)$ are discussed in Sect.~\ref{subsec:Green} and
the Appendix (see also Ref.~\cite{Meden00}).

\section{Results for the density decay}
\label{sec:results}

We next present our results for the density decay employing the three approximate approaches
discussed in the last section. As mentioned in the Introduction it is commonly believed
that Tomonaga-Luttinger liquid power laws, e.g.~the one of Eq.~(\ref{eq:friedeldecayU})
found for the decay of the Friedel oscillations, cannot be obtained by approaches based on
effective fermionic single-particle pictures. We will show that the analytical results of
the first order perturbation theory for the Green function $n^{\rm 1PT}(j)$ indicate
the TLL power law by showing a logarithmic $j$ dependence. More cannot
be expected within this approximation. The numerical results for the
non-self-consistent Hartree-Fock approximation $n^{\rm HF}(j)$ do not support that the
logarithmic behavior is elevated to a power law by the resummation inherent to the use of the Dyson
equation (\ref{eq:Dyson}). The numerical BALDA-DFT results for the density $n^{\rm DFT}(j)$ are consistent with
power-law scaling, however, with an exponent which is very close to the noninteracting value $-1$ even
for sizeable two-particle interactions.

\subsection{Perturbation theory for the Green function}
\label{subsec:Green}

To see what to expect when computing $n^{\rm 1PT}(j)$ we first expand the Tomonaga-Luttinger liquid
result Eq.~(\ref{eq:friedeldecayU}) using the leading order expression for the Tomonaga-Luttinger
liquid parameter $K$ Eq.~(\ref{eq:KsmallU}). For $k_{\rm F} \neq \pi/2$, i.e. $\nu \neq 1/2$, this
leads to
\begin{align}
  \label{eq:TLexpand}
 \!\!\!\!\!\! \left| n^{\rm TL}(j) - \nu \right| &\sim    \frac{\sin{( 2 k_{{\rm F}} j)}}{j}
 \left\{ 1 + \frac{U}{\pi v_{\rm F}} \left[ 1- \cos\left(2 k_{\rm F}\right) \right] \ln j
  \right\} \nonumber \\ 
   &+ {\mathcal O}(\left( \left[U/t\right]^2  \right) . 
\end{align}
The appearance of a $\ln j$ term in $n^{\rm 1PT}(j)$, with a prefactor which corresponds to the
negative of the leading order correction of $K$ Eq.~(\ref{eq:KsmallU}), would thus
provide an indication of Tomonaga-Luttinger liquid behavior.
In lowest order perturbation theory for the Green function we strictly expand the density
to first order in $U/t$ and thus cannot expect more, such as, e.g., a power law with a $U$ dependent
exponent.
The case of half-filling is excluded as the prefactor in front of the curly brackets on
the right hand side of Eq.~(\ref{eq:TLexpand}) would vanish. As already mentioned in
Sect.~\ref{sec:spinlferm} for a half-filled band nongeneric behavior of the density is
found \cite{Kitanine08} and from now on we exclude this from our considerations.
Results for $\nu=1/2$ are presented in Ref.~\cite{Odavic19}.

Following Ref.~\cite{Meden00} the self-energy in the basis of  the single-particle
eigenfunctions can be written as 
\begin{widetext}
\begin{align}
  \frac{(L + 1)\Sigma^{{\rm 1PT}}_{n,n'}}{U}  & =
  \!
  \Bigg\{  2 N  -   \sum\limits_{m=1}^{N}  \Big[  \cos{(k_{n} - k_{m})} + \cos{(k_{n} + k_{m})} \Big] \! \Bigg\} \delta_{n,n'}  
   - \Bigg\{  \cos{(k_{n} - k_{n'})}  -  \cos{\Bigg( \frac{k_{n} + k_{n'}}{2}\Bigg)}  \Bigg\}
  f \Bigg( \frac{\vert n - n' \vert }{2} \Bigg) \nonumber \\  
    &+ \Bigg\{  \cos{(k_{n} + k_{n'})} - \cos{\Bigg(\frac{k_{n} - k_{n'}}{2} \Bigg)}  \Bigg\}
    f \Bigg( \frac{ n + n' }{2} \Bigg)
        \label{eq:Sigman}
\end{align}
\end{widetext}
with
\begin{equation}
f(x) \! =  \!\! \left\{ \begin{array}{ll}
                           \!  1   & \quad \! \! {\rm for } \quad x \leq N \, \wedge \, x \in \mathbb{N} \\
                             \!  0 & \quad \! \! {\rm otherwise }
                           \end{array} \right.
\end{equation}
We here already neglected terms with an additional prefactor $1/L$ which are irrelevant
as we later take the  thermodynamic limit. Note that contributions from umklapp scattering,
only present for $\nu=1/2$, are suppressed.

To illustrate the effect of the two-particle interaction we first
consider the diagonal part $\Sigma^{{\rm 1PT}}_{n,n}$. 
For $L \to \infty$ it is given by
\begin{equation}
  \Sigma^{{\rm 1PT}}_{k,k} = 2 U \nu - \frac{2 U}{\pi} \sin{(\pi \nu)} \cos{k}.
  \label{eq:Sigmandia}
\end{equation}
The first addend is a $U$ dependent shift of the chemical potential. The second one can be
combined with the noninteracting single-particle dispersion $\epsilon(k) = - 2 t \cos{k}$ to 
a $U$ dependent change of the hopping $\bar t = t + \frac{U}{\pi} \sin(\pi \nu)$. In a
translational
invariant setup (periodic boundary conditions) all other matrix elements of the self-energy
vanish and on non-self-consistent Hartree-Fock level the effect of the interaction reduces to a shift of the
chemical potential and a change of the band width from $4 t$ to $4 \bar t$ (broadening of the band
for repulsive interactions $U>0$).

To obtain $n^{\rm 1PT}(j)$ we separate the noninteracting
density $n^0(j)$ and the first order correction
$\Delta n^{\rm 1PT}(j) \propto U/t$ such that $n^{\rm 1PT}(j) = n^0(j) +
\Delta n^{\rm 1PT}(j)$. Employing Eqs.~(\ref{eq:Dysonexp}) as well as
(\ref{eq:densGreen}) and performing the integral over $\omega$ we obtain for $L \to \infty$
\begin{equation}
  \Delta n^{{\rm 1PT}}(j)  = \frac{ U}{t \pi^{2}} \int_{0}^{k_{{\rm F}}}d k  \int_{k_{{\rm F}}}^{\pi} d k' \frac{\sin{(k j)} \sin{(k' j)}}{ \cos{(k')} - \cos{(k)}} \sigma^{{\rm 1PT}}_{k,k'}, \label{eq:1PTL}
\end{equation}
with
\begin{eqnarray}
  \sigma^{{\rm 1PT}}_{k,k'} \! = \!
  &-  \Bigg\{ \! \cos{(k - k')} - \cos{\Bigg( \frac{k + k'}{2} \Bigg)} \! \Bigg\} \theta \Bigg( \! k_{{\rm F}} - \frac{ k - k' }{2} \! \Bigg)  \nonumber \\
  &+   \Bigg\{ \! \cos{(k + k')} - \cos{\Bigg( \frac{k - k'}{2} \Bigg)} \! \Bigg\} \theta \Bigg(
    \! k_{{\rm F}} - \frac{ k + k' }{2} \!\Bigg) .
                                  \label{eq:smallsigmadef}
\end{eqnarray}
In the Appendix we show how to analytically evaluate the double integral for large $j$. The final result
for the leading $j$ dependence reads
\begin{equation}
  \label{eq:leadingj}
  \Delta n^{{\rm 1PT}}(j) = - \frac{\sin{( 2 k_{{\rm F}} j)}}{2 \pi j} \frac{U}{2 \pi t \sin(k_{\rm F})}
  \left[ 1- \cos\left(2 k_{\rm F}\right) \right] \ln j. 
\end{equation}  
Using $v_{\rm F} = 2t \sin k_{\rm F} $ and Eq.~(\ref{eq:n0TD}) for $n^0(j)$ the perturbative calculation
to leading order in $U/t$ agrees with our expectation from Tomonaga-Luttinger theory Eq.~(\ref{eq:TLexpand})
in the limit $j \gg 1$. Perturbation theory for the Green function is thus consistent with Tomonaga-Luttinger liquid behavior. The total density $n^{\rm 1PT}(j)$ is expected to agree with the exact one as long as the absolute value of the correction Eq. (21) is much smaller than the noninteracting density $n^0(j)$. In particular, this implies that $U/t \ln j \ll 1$ must hold.

One might hope that the resummation of higher-order terms inherent to the Dyson
equation (\ref{eq:Dyson}) will lead to the Tomonaga-Luttinger liquid power law of
Eq.~(\ref{eq:friedeldecayU}) for  $n^{\rm HF}(j)$
with the correct leading order (in $U/t$) exponent instead of the logarithmic behavior found for
$n^{\rm 1PT}(j)$. This will be investigated next.  

\subsection{The non-self-consistent Hartree-Fock approximation}
\label{subsec:self}

Due to the nontrivial spatial dependence of the self-energy Eqs.~(\ref{eq:SigmaHFdia}) and
(\ref{eq:SigmaHFoffdia}) we did not succeed in analytically performing the inversion inherent
to Eq.~(\ref{eq:Dyson}).
All non-self-consistent Hartree-Fock results  $n^{\rm HF}(j)$ shown in this section were thus
obtained by inserting the self-energy matrix elements Eqs.~(\ref{eq:SigmaHFdia}) and
(\ref{eq:SigmaHFoffdia}) into Eq.~(\ref{eq:Dyson}), determining the diagonal part of the inverse
by the ${\mathcal O}(L)$ algorithm of Ref.~\cite{Andergassen04}, and numerically performing the
integral Eq.~(\ref{eq:densGreen}) (implemented as the solution of a differential equation).

\begin{figure}
\resizebox{\columnwidth}{!}{%
  \includegraphics{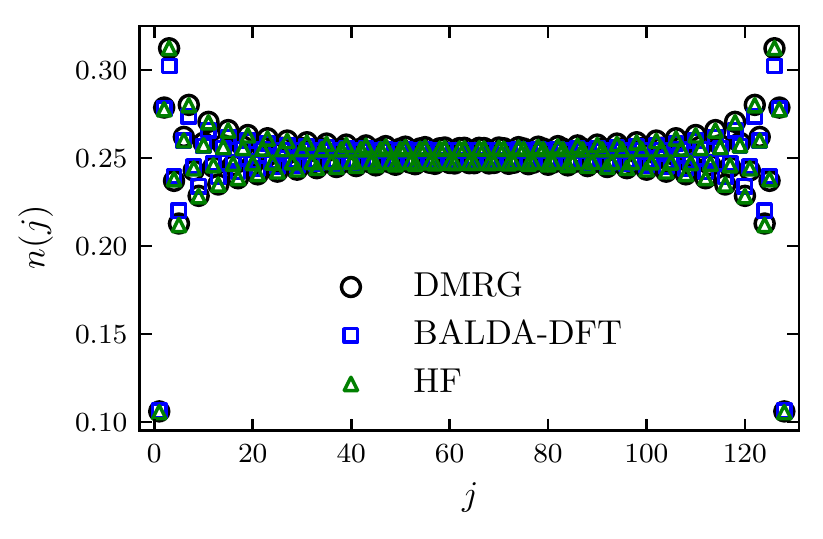}
}
\caption{Density profile $n(j)$ for $L = 128$ sites and interaction strength $U/t = 0.5$ at
  quarter-filling $\nu = 1/4$. Results obtained within the non-self-consistent Hartree-Fock approximation
  and the BALDA-DFT are compared to ``numerically exact'' ones computed using DMRG. The DMRG data were provided
  by C.~Karrasch.}
\label{fig1}       
\end{figure}

In Fig.~\ref{fig1} we compare   $n^{\rm HF}(j)$ (and in addition $n^{\rm DFT}(j)$; see Sect.~\ref{subsec:DFT})
with highly accurate (``numerical exact'') results obtained using the density-matrix
renormalization group (DMRG). This numerical approach can be used for systems of up to $10^3$ lattice sites.
In the figure we show data for $L=128$, $\nu=1/4$, and $U/t=0.5$. The overall agreement is acceptable. The
data clearly show the $2k_{\rm F}$ periodicity. An analysis
of the decay in the light of the asymptotic Tomonaga-Luttinger
liquid power law Eq.~(\ref{eq:friedeldecayU}) is meaningless as the overlap of the oscillations
originating from the two boundaries of the chain prevents that the asymptotic behavior develops
for such small systems. Therefore, larger systems have to be studied.

Using the ${\mathcal O}(L)$ algorithm discussed in Sect.~\ref{sec:PT} we can straightforwardly
compute $n^{\rm HF}(j)$ for systems of up to $L=10^6$ sites. We believe that for a faithful
and unbiased search for power-law scaling a corresponding fit of the numerical data is not
sufficient. In particular, using perturbation theory we are bound to small interactions, for
which the exact exponent is very close to the noninteracting value $-1$; see Eq.~(\ref{eq:KsmallU}).
In this case a power law might be barely distinguishable from the leading logarithmic behavior
analytically found in perturbation theory for the Green function Eq.~(\ref{eq:leadingj}).
We will thus perform a more stringent
analysis of the data. To this end we focus on the upper envelope of the decaying data (compare
Fig.~\ref{fig1}), that is the $j$ for which a local maximum is taken. They have a mutual
distance of $\nu^{-1}$. We then compute centered logarithmic differences
\begin{equation}
  \label{eq:logderiv}
  \alpha(j) = \frac{\ln\left[\tilde n\left(j+\nu^{-1}\right)\right] - \ln\left[\tilde n\left(j-\nu^{-1}\right)\right]}{\ln
  \left(j+\nu^{-1}\right) - \ln \left(j-\nu^{-1}\right)},
\end{equation}
with $\tilde n(j) = n(j) - \nu$ ,
which should approach a constant value (the value of the exponent) for sufficiently large $j$
if the density decays according to a power law.
In contrast to a power-law fit the logarithmic differences directly indicate any systematic
deviation from power-law behavior. In addition, we compute  the following semi-logarithmic centered
differences
\begin{equation}
  \label{eq:semilogderiv}
  \beta(j) = 2 \pi
\frac{\left(j+\nu^{-1}\right) \tilde n\left(j+\nu^{-1}\right) - \left(j-\nu^{-1}\right) \tilde n\left(j-\nu^{-1}\right)}{\ln
  \left(j+\nu^{-1}\right) - \ln \left(j-\nu^{-1}\right)}.  
\end{equation} 
If also the  non-self-consistent Hartree-Fock data only show the logarithmic correction
Eq.~(\ref{eq:leadingj}) instead of a resummed power law, $\beta(j)$ should display a plateau
at the value $\frac{U}{2 \pi t \sin(k_{\rm F})}
\left[ 1- \cos\left(2 k_{\rm F}\right) \right]$. We then compare $\alpha(j)$ and $\beta(j)$ for a
given parameter set to judge which of the two is more plateau-like and thus to judge if the data
are more consistent with a power law or the logarithmic behavior Eq.~(\ref{eq:leadingj}).
Note that taking the logarithmic differences Eq.~(\ref{eq:logderiv}) or semi-logarithmic
ones Eq.~(\ref{eq:semilogderiv})
significantly enhances any small numerical error in $n(j)$.

\begin{figure}
\resizebox{\columnwidth}{!}{%
  \includegraphics{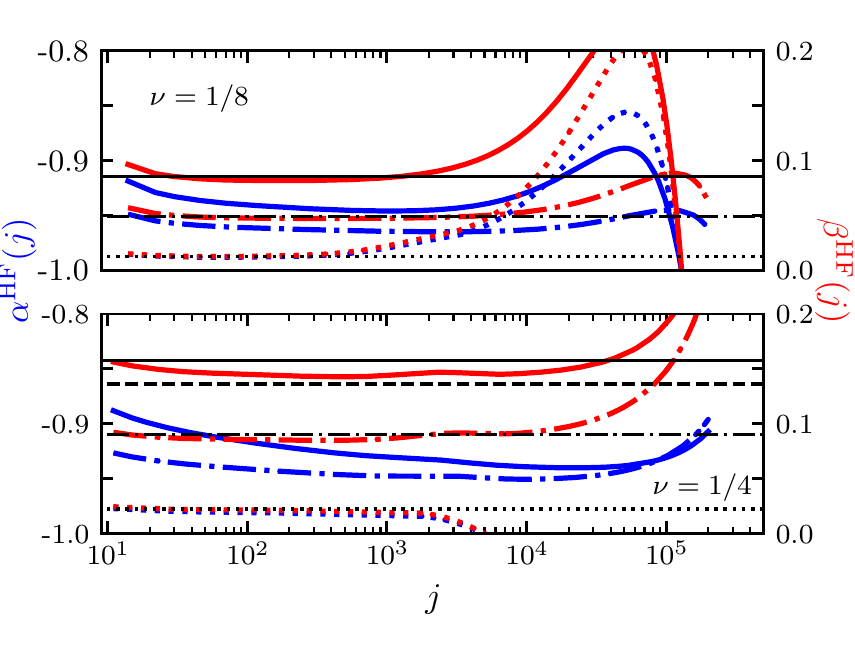}
}
\caption{The apparent exponent $\alpha(j)$ Eq.~(\ref{eq:logderiv}) (blue) and
  the apparent prefactor of $\ln j$ Eq.~(\ref{eq:semilogderiv}) (red) of the decay of the
  density oscillations within the non-self-consistent Hartree-Fock approximation for a system with
  two open boundaries. The parameters are $\nu=1/8$ (upper panel) and $\nu=1/4$ (lower panel),
  $U/t=0.1$ (dotted), $U/t=0.4$ (dashed-dotted), and $U/t=0.7$ (solid). The horizontal lines indicate
  the corresponding prefactor of the $\ln j$ term from first order perturbation theory for the Green function
  Eq.~(\ref{eq:leadingj}). The horizontal dashed line shows this value with $t$ replaced by $\bar t$
  (only shown for $\nu=1/4$ and $U/t=0.7$).}
\label{fig2}       
\resizebox{\columnwidth}{!}{%
  \includegraphics{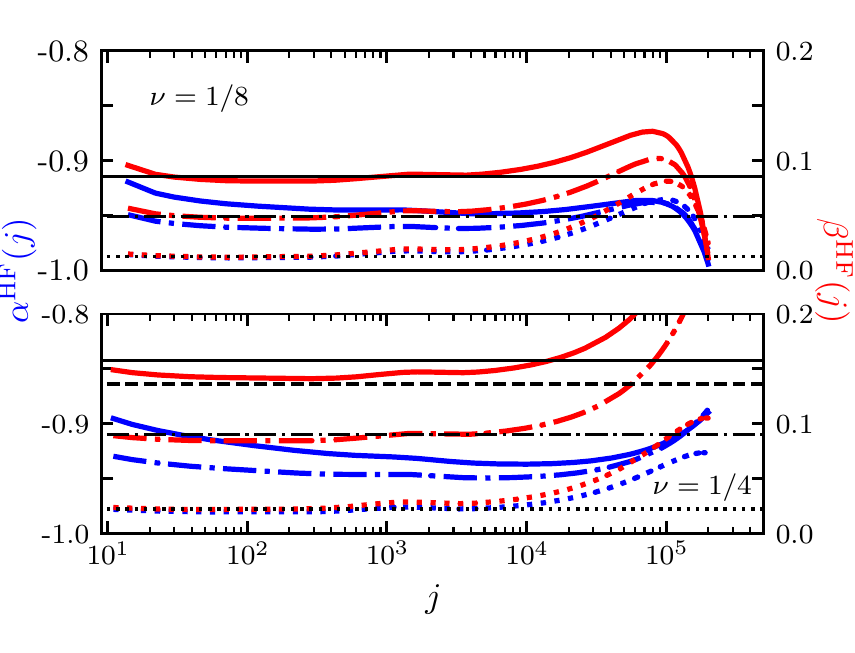}
}
\caption{The same as in Fig.~\ref{fig2}, but for a chain with one open boundary, which at $j=L$ is
  adiabatically connected to a semi-infinite noninteracting lead.}
  \label{fig3}       

\end{figure}

In Fig.~\ref{fig2} we show $\alpha^{\rm HF}(j)$ and $\beta^{\rm HF}(j)$ for $L=2^{20}$,
the two fillings $\nu=1/4$ and $\nu=1/8$ as
well as three interactions $U/t=0.1$, $U/t=0.4$, and $U/t=0.7$.
While for small $U/t$ the numerical non-self-consistent Hartree-Fock data
are consistent with both the power law and the logarithmic behavior Eq.~(\ref{eq:leadingj}), for
larger $U/t$, $\beta^{\rm HF}(j)$ is more plateau-like as compared to
$\alpha^{\rm HF}(j)$. Furthermore,
the value of the plateau of $\beta^{\rm HF}(j)$ is close to $\frac{U}{2 \pi t \sin(k_{\rm F})}
\left[ 1- \cos\left(2 k_{\rm F}\right) \right]$ which is shown as the horizontal
lines in Fig.~\ref{fig2}. However, due to the effect of the right boundary, at larger $j$ a
deviation from the plateau is found already for $j \ll L$.

For $\nu=1/4$ and $U/t=0.7$ a deviation between the plateau value of the data and the expectation
(solid horizontal line) from first order perturbation theory is found. This is due to higher order
corrections (in $U/t$) appearing in the non-self-consistent Hartree-Fock approximation. For larger
interactions those become sizable. One such correction originates from the changed band width
$4t \to 4 \bar t$ as discussed in Sect.~\ref{subsec:Green}. In fact, in the non-self-consistent
Hartree-Fock approximation $t$ is replaced by $\bar t$ at any instance the hopping amplitude appears.
For this reason the horizontal dashed line (only shown for $\nu=1/4$ and $U/t=0.7$) which indicates $\frac{U}{2 \pi \bar t \sin(k_{\rm F})}
\left[ 1- \cos\left(2 k_{\rm F}\right) \right]$  fits better
to the data. 

To further investigate  $n^{\rm HF}(j)$ we suppress the effect of the right boundary by
adiabatically connecting the interacting chain to a semi-infinite noninteracting tight-binding chain
at site $j=L$. This way the spatial region $j \approx L$ does not act as a source of any
(significant) oscillations in the self-energy and thus not as a source of another decaying oscillation
in the density. The technical details how to achieve this are described in Ref.~\cite{Andergassen04}.
In particular, the interaction has to be turned off smoothly over a sufficiently large spatial regime close to
$j=L$. Figure \ref{fig3} shows
$\alpha^{\rm HF}(j)$ and $\beta^{\rm HF}(j)$ obtained this way for the same parameters as in Fig.~\ref{fig2}.
As expected, for most parameter sets the plateau in $\beta^{\rm HF}(j)$ extends towards larger
$j$ if the oscillations originating from $j \approx L$ are suppressed. We generically gain
between one half and one order of magnitude; compare Figs.~\ref{fig3} and \ref{fig2}.
However, our conclusions drawn are the same as the ones from  the setup with two open boundaries. 
Even with a noninteracting lead connected adiabatically deviations from the plateau value are
found already at $j < L$. The information about
the finiteness of the interacting part of the chain is still encoded in the data.

Based on these results we conclude that the resummation inherent to the Dyson equation (\ref{eq:Dyson})
does not lead to a resummation of the logarithmic behavior Eq.~(\ref{eq:leadingj}) of first
order perturbation theory to a power law. This has to be contrasted to the frequency dependence
of the local single-particle spectral function in which this resummation was shown earlier
\cite{Andergassen04}.

We note in passing, that a self-consistent Hartree-Fock approximation leads to Friedel oscillations
with an amplitude which is much larger than the one found using DMRG.  Increasing the system size $L$
a nondecaying density oscillation appears to develop; see Ref.~\cite{Sch} and \cite{Odavic19}. The
self-consistency seemingly triggers a spurious charge-density wave instability. This is not surprising
as the 1d system is highly susceptible towards $2 k_{\rm F}$ instabilities. The self-consistent
Hartree-Fock approximation is thus an inappropriate approach to study the problem at hand. 

In our search for an approximate method, which is based on an effective
fermionic single-particle picture, to capture the Tomonaga-Luttinger liquid
power law Eq.~(\ref{eq:friedeldecayU}), in the next section we use BALDA-DFT. As it is usually the case in a DFT approach the regime of validity given a certain exchange-correlation functional is not obvious a priori. In fact, this is one of our motivations to study the density within BALDA-DFT.

\subsection{BALDA-DFT}
\label{subsec:DFT}

We finally investigate whether or not the LDA-DFT with an exchange-correlation functional determined
from the exact Bethe ansatz solution of the homogeneous system is able to produce the Tomonaga-Luttinger
liquid power-law decay of the Friedel density oscillations.

Figure \ref{fig1} shows  $n^{\rm DFT}(j)$ in comparison to $n^{\rm DMRG}(j)$ and $n^{\rm HF}(j)$
for a small system with $L=128$, $\nu=1/4$, and $U/t=0.5$. The BALDA-DFT results are very close to the
ones obtained within the non-self-consistent Hartree-Fock approximation.

\begin{figure}
\resizebox{\columnwidth}{!}{%
  \includegraphics{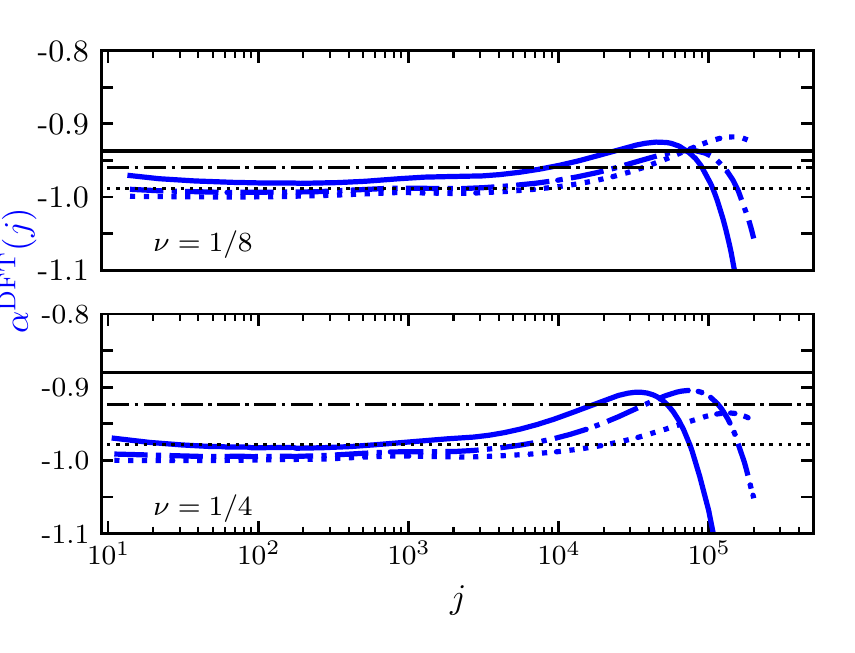}
}
\caption{The apparent exponent $\alpha(j)$ Eq.~(\ref{eq:logderiv}) of the decay of the
  density oscillations within the BALDA-DFT approximation for a chain with one open boundary, which at $j=L$ is
  adiabatically connected to a semi-infinite noninteracting lead. The parameters are the same as in Fig.~\ref{fig2}.
  Here, the horizontal lines indicate the exact exponent $-K(\nu,U/t)$, with $U/t=0.1$ (dotted), $U/t=0.4$ (dashed-dotted), and $U/t=0.7$ (solid).}
\label{fig4}       
\end{figure}

In Fig.~\ref{fig4} we again consider a larger chain, $L=2^{20}$ and show the logarithmic
derivative (the apparent exponent)
$\alpha^{\rm DFT}(j)$ for the same parameters as in Fig.~\ref{fig2}. Here, we directly study the chain
with one open boundary which at $j=L$ is adiabatically connected to a semi-infinite noninteracting
lead, a setup which turned out to be advantageous for the analysis of the non-self-consistent
Hartree-Fock data.  The data show a plateau at intermediate $j$ and are therefore consistent with
power-law behavior, however, with an exponent which is far off from the exact one
$-K(\nu,U/t)$ (horizontal lines). Based on these results one is tempted to
conclude that the BALDA-DFT exponent agrees with the noninteracting value $-1$.

Obviously, the BALDA-DFT does not give a satisfying description of the Tomonaga-Luttinger liquid
power-law scaling of the decay of the density away from an open boundary towards the bulk
value $\nu$. It, however, does not lead to a spurious charge-density wave instability as found
in self-consistent Hartree-Fock. We emphasize that
the failure of BALDA-DFT to correctly describe the de-
cay of the Friedel oscillations is entirely due to the use of
the Bethe ansatz local density approximation. The ex-
act functional, which is unknown, would reproduce the
many-body density and hence describe the decay of the
oscillations correctly. 

\section{Conclusion}
\label{sec:conclusion}

In this paper we have provided strong numerical evidence that neither the non-self-consistent
Hartree-Fock approximation nor a density functional theory approach within the Bethe-ansatz local density
approximation are able to capture the Tomonaga-Luttinger liquid power-law
decay of the Friedel density oscillations off an open boundary. We focused on the 1d lattice
model of spinless fermions with nearest-neighbor hopping and (short-range) nearest-neighbor
two-particle interaction. 

As expected, first order many-body perturbation theory of the Green function (in the two-particle
interaction) shows a logarithmic dependence of the density $n(j)$ on the position which is in
accordance with Tomonaga-Luttinger liquid behavior. The numerical data of the non-self-consistent
Hartree-Fock approximation indicate that the resummation of higher order terms inherent to the
use of the Dyson equation within this approach does not elevate this logarithmic behavior
to a power law. The data for the density are rather consistent with $\ln j$ behavior. This
has to be contrasted to another observable, the local spectral function close to an open
boundary as a function of frequency, for which such a resummation was observed when going
from first order perturbation theory for the Green function to the non-self-consistent
Hartree-Fock approximation \cite{Meden00}. We briefly mentioned that the self-consistent
Hartree-Fock approximation is prone to a spurious $2 k_{\rm F}$ charge-density wave
instability \cite{Sch,Odavic19}.

For the 1d (spinful) Hubbard model the decay of the Friedel density oscillations off
an impurity was earlier investigated using BALDA-DFT \cite{Lima03}. Even before
discussing our BALDA-DFT data for the density, we provided arguments which clearly
indicate that the results of Ref.~\cite{Lima03} were obtained for system sizes which
are way to small to allow for a meaningful search for the asymptotic Tomonaga-Luttinger
liquid power-law decay. Studying systems of up to $10^6$ lattice sites for a spinless model
with open boundaries we are in a position to investigate if BALDA-DFT captures this
power law. Our numerical data are consistent with a power-law decay of the density
oscillations towards the bulk density, however,
with the noninteracting exponent $-1$ instead of the interaction and filling dependent
one $-K(\nu,U/t)$. We thus conclude that BALDA-DFT does not capture the
Tomonaga-Luttinger liquid characteristics of $n(j)$. This result is in accordance
with the conclusion reached in Ref.~\cite{Schenk08}, in which the same model as
studied here was investigated. In this paper observables other than the density were
computed for systems of a few hundred lattice sites. We reiterate that besides
the ground-state energy the density is the observable most directly accessible
in a DFT approach. As such one can hope that future
improvements to the BALDA functional will be able to
describe the decay of the density oscillations correctly.

\section{Acknowledgments}
  This work was supported by the Deutsche Forschungsgemeinschaft via RTG 1995. N.H.
  acknowledges additional funding from an Emmy-Noether grant of the Deutsche Forschungsgemeinschaft.
  We thank C.~Karrasch for providing the DMRG data of Fig.~\ref{fig1}, M.~Pletyukhov for
  discussions on the asymptotic analysis presented in the Appendix, and Nicolai Kitanine
  for discussions on the Bethe ansatz solution.
  
\appendix

\section{Details on the perturbation theory for the Green function}
\label{appendix}

In this Appendix, we present details on the asymptotic analysis ($j \rightarrow \infty$) for
the Fourier type integrals in Eq.\ (\ref{eq:1PTL}). We separately discuss integrals over
rectangular and triangular domains, which correspond to the terms that appear in the first
and second line of Eq.\ (\ref{eq:smallsigmadef}), respectively.

\subsection{Integrals over a rectangular domain}

\begin{figure}[ht]
\resizebox{\columnwidth}{!}{%
  \includegraphics{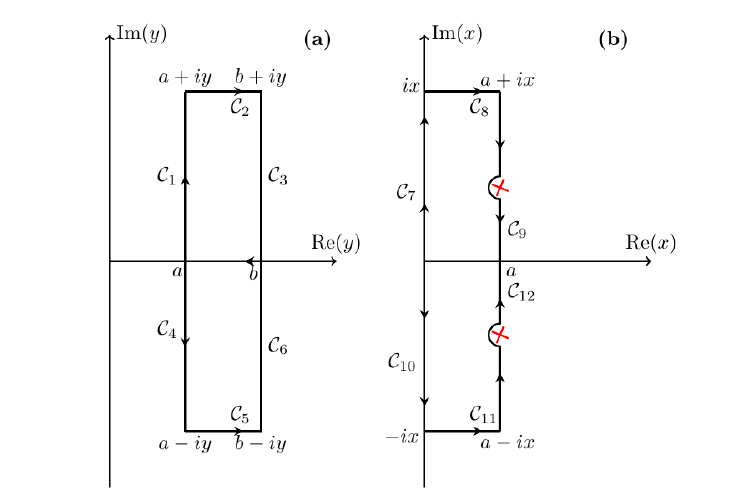}
}
\caption{(a) Integration contour of the inner integral. (b)  Integration contour of the outer integral.
  The poles are indicated by the red crosses.}
\label{appendix_fig1}       
\end{figure}

Integrals that appear take the following form (for $ 0 < a < b \in \mathbb{R}$) 
\begin{equation}
I^{{\rm R}} = \int\limits_{0}^{a} \mathrm{d} x \int\limits_{a}^{b} \mathrm{d} y \frac{\sin{(x j)} \sin{(y j)}}{\cos{(y)} - \cos{(x)}} f (x, y), \label{eq:appendix_1}
\end{equation}
where $f(x,y)$ is assumed to be analytic, and symmetric under the exchange of the arguments for
the following analysis to apply. These conditions are satisfied by the terms in the first
line of Eq.\ (\ref{eq:smallsigmadef}). The integrand of Eq.~(\ref{eq:appendix_1}) is singular at
$x = y = a$. Hence, integration by parts cannot be employed to extract the asymptotics.
Therefore, we rewrite the integral as
\begin{equation}
  I^{{\rm R}} = \int\limits_{0}^{a} \mathrm{d} x \int\limits_{a}^{b} \mathrm{d} y  \Big( e^{i x j} - e^{-i x j} \Big) \Big( e^{ i y j} - e^{-i y j} \Big) g(x,y), \label{eq:appendix_2}
\end{equation}
with $g (x,y) = - \frac{1}{4} \frac{f (x, y)}{\cos{(y)} - \cos{(x)}}$, and proceed by
using the method of steepest descent. We deform the integration contour $\mathcal{C}$,
which runs from $a$ to $b$ along the real $y$ axis as depicted in Fig.\ \ref{appendix_fig1} (a),
into $\mathcal{C}_{1} \cup \mathcal{C}_2 \cup \mathcal{C}_3 \cup \mathcal{C}_4 \cup
\mathcal{C}_5 \cup \mathcal{C}_6$. The integral now reads
\begin{equation}
I^{{\rm R}} = \int\limits_{0}^{\infty} \mathrm{d} (i y) e^{-j y} \int\limits_{0}^{a} \mathrm{d} x h(x) \Big( e^{i x j } - e^{- i x j} \Big), \label{eq:appendix_3}
\end{equation}  
with $h (x) = e^{i j a} g (x, a + i y) - e^{i j b} g (x, b + i y)+ e^{- i j a} g(x ,a - iy) - e^{-i j b}
g (x, b - i y) $. We apply the same steps to the integral over $x$ and deform the contour into three
line segments for each of the two terms as before for the integration over $y$.

This time, however,
two poles are located on the contour, which arise from the singularity. They
were shifted away from
the real axis as depicted in Fig.\ \ref{appendix_fig1} (b). Following this second contour deformation the integral reads
\begin{widetext}
\begin{align}
&  I^{{\rm R}} = \int\limits_{0}^{\infty} \mathrm{d} (i y) e^{- y j} \Bigg\{
i \pi {\rm Res} \Big[ h (a+ i x) e^{i j (a + i x)}, x \rightarrow y \Big]  + i \pi {\rm Res} \Big[ h (a - i x) e^{-i j (a - i x)}, x \rightarrow y \Big]  \nonumber \\
&+ \int\limits_{0}^{\infty} \mathrm{d} (i x) e^{- x j} \Big( 2 {\rm Re} \left[h(ix)\right] - e^{i j a} h (a + i x) - e^{- i j a} h (a - i x) \Big) \Bigg\},\label{eq:appendix_4}
\end{align}
\end{widetext}
where ${\rm Res}$ denotes the residue. The poles only contribute with half their residue as they lie on
the contour. The integrals in the second line of Eq.\ (\ref{eq:appendix_4}) are either zero, because
the domain of integration is symmetric under reflection with respect to the axis $x = y$, cancel each
other under the exchange of arguments, or yield sub-leading contributions $ \sim j^{-2}$. The integrals
that produce the sub-leading contributions do not contain the pole and can be computed using integration
by parts. The leading order contribution to the integral originates from the residues, which when
evaluated, give
\begin{align}
  I^{{\rm R}} & =  \frac{\pi}{4} \int\limits_{0}^{\infty} \mathrm{d} y  e^{ - 2 j y} \Bigg(
                   \frac{f(a + i y,a + i y)}{\sin{(a + i y)}} e^{2 i j a}  \nonumber \\
&+ \frac{f(a- i y,a - i y)}{\sin{(a - i y)}} e^{-2 i j a}  \Bigg) + \mathcal{O} \big( j^{-2} \big).
\end{align}
After substituting $y'=jy$ and taking the limit $j \rightarrow \infty$ we obtain the leading order 
asymptotic contribution as
\begin{equation}
  I^{{\rm R}} \approx  \frac{\pi}{4 j} \frac{f (a, a)}{\sin{(a)}} \cos{(2 j a)}
  \label{eq:asym1}.
\end{equation}

\subsection{Integrals over a triangular domain}

The integrals we encounter on the triangular domains are of the following form
\begin{equation}
  I^{{\rm T}} = \int\limits_{0}^{a} \mathrm{d} x \int\limits_{a}^{2 a - x} \mathrm{d} y
  \frac{\sin{(x j)} \sin{(y j)}}{\cos{(y)} - \cos{(x)}} f (x,y).
\end{equation}
The same constraints on $f(x,y)$ as for the integrals on rectangular domains hold.
Rewriting the integrand in terms of exponential functions and using the
transformation 
as $(x,y) \mapsto (a - x, a + y) $ we obtain
\begin{align}
 I^{{\rm T}} &= \int\limits_{0}^{a} \mathrm{d} x \int\limits_{0}^{x} \mathrm{d} y  g (a - x, a + y)  
\times \Big( e^{i j (a - x)} - e^{- i j (a - x)} \Big) \times \nonumber \\
& \Big( e^{i j (a + y)} - e^{-i j (a + y)} \Big).
\end{align}
The transformation shifts the pole from $(a,a)$ to $(0,0)$ and simplifies the computation by
fixing the singularity. Using contour extensions for the integral over $y$ we arrive at
\begin{align}
\tilde I^{{\rm T}}  &= e^{i j a} \int\limits_{0}^{a} \mathrm{d} x \int\limits_{0}^{\infty} \mathrm{d} (i y) e^{- j y} 
\nonumber \\
&\times \Bigg\{ \Big( e^{i j (a - x)} - e^{-i j (a - x)} \Big)   g (a - x, a + iy)  \nonumber \\ 
&- \Big( e^{i j a} - e^{-i j (a - 2 x)} \Big) g(a - x, a + x + iy)  \Bigg\},
\end{align}
where $I^{{\rm T}} = 2 {\rm Re} \tilde I^{{\rm T}}$. $\tilde I^{{\rm T}}$ contains four integrals.
Two of those can be evaluated to yield contributions
to sub-leading order $\sim j^{-2}$. One contains the pole and computing the residue, as in the case of rectangular
integration domains, we obtain to leading order
\begin{equation}
\tilde  I^{{\rm T}}_{\rm pole} \approx \frac{\pi}{8 j \sin{a}} f (a, a) e^{2 i j a}. \label{eq:appendix_5}
\end{equation}
The remaining term reads
\begin{equation}
  \frac{e^{2 i j a}}{4}  \int\limits_{0}^{\infty} \mathrm{d} (i y) e^{- j y}
  \int\limits_{0}^{a} \mathrm{d} x \frac{f(a - x,a + x + i y)}{\cos{(a + x + iy)}-\cos{(a - x)}}.
\end{equation}
At this point we need to specify the function $f$ to be able to proceed. As an example, we consider
$f(x,y) = \cos{(x + y)}$; see Eq.\ (\ref{eq:smallsigmadef}).  The integral over $x$ is $j$
independent and can evaluated straightforwardly. It yields
\begin{align}
  - \frac{i e^{2 i j a}}{8} &\int\limits_{0}^{\infty} \mathrm{d} y e^{- j y} \frac{\cos{(2 a + i y)}}{\sin{\Big( a + \frac{i y}{2} \Big)}} \nonumber \\
&\times\ln{\Bigg[ \tan{\Big(a/2 + i y/4 \Big)} \cot{\Big( i y/4 \Big)} \Bigg]}.
\end{align}
Taking the limit $j \rightarrow \infty$ and using that 
\begin{equation}
\int\limits_{0}^{\infty} e^{- x j} \ln{(x)} d x = - \frac{\gamma + \ln{(j)}}{j},
\end{equation}
where $\gamma \approx 0.577...$ is the Euler-Mascheroni constant, we obtain
a logarithmic contribution to the integral. Combining this with the pole
contribution of Eq.\ (\ref{eq:appendix_5}) we finally obtain for the leading order asymptotics
\begin{align}
  I^{{\rm T}} &\approx \frac{\sin{(2 j a)} \cos{(2 a)}}{ 4 j \sin{(a)}}  \Big\{ \ln{j}+ \gamma + \ln{4 \tan{( a/2)}} \Big\} \nonumber \\
  &+ \frac{\cos{(2 j a)} \pi \cos{(2 a)}}{ 8 j \sin{(a)}}.
                 \label{eq:asym2}
\end{align}
To verify that the analytical expressions Eqs.~(\ref{eq:asym1}) and (\ref{eq:asym2})
for the asymptotic behavior of $I^{{\rm R}}$ and $I^{{\rm T}}$ are
indeed correct, we performed numerical integrations for large $j$'s
and found agreement. Due to the divergence of the integrand, a particular set of
transformations to the integrals needs to be applied before it can reliably be computed
numerically. For more details, see Ref.\ \cite{Odavic19}.

\section{Authors contributions}
All the authors were involved in the preparation of the manuscript.
All the authors have read and approved the final manuscript.
%
%

\end{document}